\documentclass[12pt]{iopart}
\usepackage{iopams} 
\usepackage{graphicx}

\newcommand{\bra}[1]{\langle #1 |} 
\newcommand{\ket}[1]{| #1 \rangle} 
\newcommand{\braket}[2]{\langle #1 | #2 \rangle} 

\begin{document}

\title[]{Nuclear wave function interference in single-molecule electron transport
}

\author{Maarten R. Wegewijs
  \footnote[3]{To whom correspondence should be addressed (wegewijs@physik.rwth-aachen.de)}
  \ and Katja C. Nowack }

\address{ Institut f\"ur Theoretische
   Physik - Lehrstuhl A , RWTH Aachen , 52056 Aachen , Germany }

\begin{abstract}
It is demonstrated that non-equilibrium vibrational effects are
enhanced in molecular devices for which  the effective potential for
vibrations is sensitive to the charge state of the device.
We calculate the electron tunneling current through a molecule
accounting for the two simplest qualitative effects of the charging
on the nuclear potential for vibrational motion:
 a shift (change in the equilibrium position) and
 a distortion (change in the vibrational frequency).
The distortion has two important effects: firstly, it breaks the
symmetry between the excitation spectra of the two charge states.
This gives rise to new transport effects which
map out changes in the current-induced non-equilibrium vibrational
distribution with increasing bias voltage.
Secondly, the distortion modifies the Franck-Condon factors for
electron tunneling. Together with the spectral asymmetry this gives
rise to pronounced nuclear wave function interference effects
 on the electron transport.
For instance nuclear-parity forbidden transitions lead to differential
conductance {\em anti-resonances}, which are stronger than those due
to allowed transitions.
For special distortion and shift combinations a {\em coherent suppression}
of transport beyond a bias voltage threshold is possible.
\end{abstract}


\maketitle
\section{Introduction}
The question how quantized vibrational modes affect the
electron transport through a single molecule has recently attracted a
lot of interest, both
 experimentally~\cite{Park00,Park02,Pasupathy04,Yu04kondo,Yu04ndc,Yu04c60,LeRoy04} and
 theoretically~\cite{Boese01,McCarthy03,Braig03a,Braig04b,Mitra04b,Koch04b,Koch04c,Cizek04,Koch05a,Cornaglia04,Cornaglia05}.
For one, charging a molecule in a junction may induce a shuttling of
the center of mass~\cite{Boese01,McCarthy03}
through the interaction with the electrodes.
It may also change the internal configuration of the molecule by
changing bond lengths or angles~\cite{Cizek04}.
In most of the cited works Coulomb charging effects on the molecule
played an important role.
Previously we demonstrated that this is even more so when multiple
orbitals couple asymmetrically to a vibrational mode and start
competing~\cite{Nowack05}, cf. also~\cite{Kaat05}.
Up to now theoretical models for transport have assumed the {\em frequency}
of the oscillation to be independent of the charge state of the
molecule.
In this paper we investigate the effects of a {\em distortion} in
addition to a shift of the nuclear potential with respect to some
coordinate internal to the molecule.
We focus on the simple case of a single orbital (i.e. two relevant charge
states).
Situations may occur where the shift of the potential is small
for symmetry reasons and the distortion plays a prominent role.
For instance for a hindered rotation, charging of the molecule may
lower the rotational barrier without shifting the potential minimum.
The resulting change in frequency may be large as cases are known
where the barrier is even inverted (i.e. potential minima and maxima interchange
roles~\cite{Cizek04}).
In general, an extreme flattening of the potential surface may also
lower the energy of a dissociative continuum of states of the nuclear
motion. This will not be considered here, see however~\cite{Koch05a}.
We study the simplest transport model possible incorporating
distortion effects for bound nuclear motion 
together with the usual shift of the nuclear potential minima.
We emphasize that this is the most general harmonic approximation to
the nuclear potentials of a pair of charge states involved in
transport.
This model is readily extended to more detailed potential surfaces
 adapted to specific situations.
We focus mainly on low energy excitations for which anharmonic effects
lead only to quantitative corrections.
\\
An important property of harmonic potentials is their spatial
inversion symmetry with respect to the minimum.
If the shift is sufficiently small relative to the distortion,
the nuclear wave function parity is quasi-conserved.
This imposes a quasi-selection rule on the Franck-Condon (FC) factors
which determine the rates for the electron tunneling.
This leads to effects in the {\em intensity} of the non-linear
conductance resonances which are related to destructive interference
of the nuclear wave functions.
This is to be contrasted to the blockade~\cite{Mitra04b,Koch04b}
and NDC effects~\cite{Nowack05} discussed previously, which
basically follow from classical features of the nuclear motion
even though the discrete vibrational excitations are of quantum
signature.
Quantum effects in the intensities of the
conductance resonances may also occur for moderate shifts of the
distorted potentials.
The FC-factors of a single low-lying vibrational excitation may become
strongly asymmetric due to constructive {\em and} destructive
interference, thereby strongly increasing its lifetime.
In a large region of applied voltages this leads to a bias driven population
inversion of the vibrational states
 which suppresses the electronic transport.
This is only possible when the zero-point motions (ZPM) of the
vibration in different charge states are sufficiently different due to
a distortion.
\\
The paper is organized as follows. In Sect.~\ref{sec:model} we
formulate the transport model and discuss how the distortion affects
the non-equilibrium transport through the quasi-classical and quantum
features of the FC-factors (Sect.~\ref{sec:fc}).
Relaxation of the vibrational excitations is incorporated (Sect.~\ref{sec:relax}) in order to
identify non-equilibrium effects by comparison.
In Sect.~\ref{sec:results} we discuss the results and transport
mechanisms for weak and strong distortion. We conclude with a
discussion in Sect.~\ref{sec:discuss}.
\section{Model
\label{sec:model}
}
\begin{figure}
\begin{center}
   \includegraphics[scale=0.45]{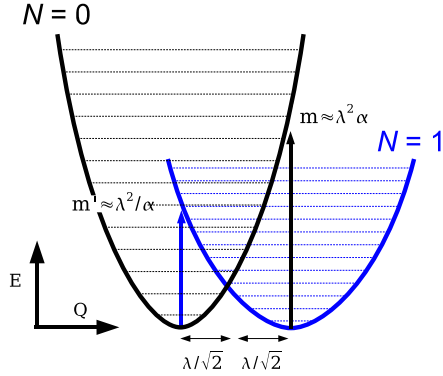}
   \caption{\label{fig:pot}
     Schematic effective nuclear potential (cf. Eq.~(\ref{eq:HN})) for
     the neutral ($N=0$, black) and charged ($N=1$, blue) electronic state of
     the molecule. The large charging energy has been subtracted since
      near the charge degeneracy point only changes in the
     {\em vibrational} energy are of importance for the differential
      conductance.
     The potentials are plotted as function of the normalized
     coordinate $Q$ (see text)
     and are relatively shifted by $\sqrt{2}\lambda > 0$ and
     distorted by a factor $\alpha^2=\omega_0 / \omega_1 > 1$.
     The vertical arrows indicate the change in the classical elastic
     energy
     when charging (blue) and de-charging (black) the molecule,
     respectively, while maintaining the nuclei at the equilibrium
     position of the initial charge state.
   }
\end{center}
\end{figure}
We consider a molecule which is weakly tunnel-coupled to a left (L),
and right (R) metallic electrode and capacitively coupled to a gate
electrode.
The transport is assumed to be dominated
by a single orbital on the molecule
and double occupancy of this orbitals prohibited by strong Coulomb blockade.
For simplicity we disregard effects of electron spin degeneracy (the minor
quantitative corrections are trivial to include)
so we can consider spinless non-interacting electrons.
Nuclear motion internal to the molecule is taken into account through a single vibrational
mode which couples both linearly and quadratically to the charge on
the molecule.
As mentioned above, quadratic coupling may certainly compete with
linear effects.
The Hamiltonian $H=H_{\rm{M}}+\sum_r H_{r}+H_{\rm{T}}$ incorporates
the molecule ($\rm{M}$), reservoirs $r=L,R$ and the tunneling between
molecule and electrodes ($\rm{T}$):
 \begin{eqnarray}
  H_{\rm{M}} & = &
  \sum_{N=0,1} \ket{N}\bra{N} H_N
  ,
  \label{eq:HM}\\
  H_N        & = &
     \frac{p^2}{2 M}
   + \frac{M}{2} \omega_N^2 \left( x \pm \delta x / 2 \right)^2
   ,
  \label{eq:HN_bare}
  \\  
  H_{\rm{e}} & = &
  \sum_{r=R,L} \sum_{k}\epsilon_{k r}a_{k r}^{\dag}a_{k r}
  ,
  \label{eq:He}
  \\
  H_{\rm{T}} & = & \sum_{r=R,L} \sum_{k} t_r a_{k r}^{\dag}\ket{0}\bra{1} + h.c.
  ,
  \label{eq:HT}
 \end{eqnarray}
where $\pm$ corresponds to $\ket{N}$ denoting the
neutral ($N=0$) and charged ($N=1$) electronic state of the molecule
respectively.
We employ units $\hbar=k_{B}=e=1$.
The electrodes $r=L,R$ are described by the non-interacting Hamiltonian
$H_{\rm{e}}$.
They are maintained at fixed temperature $T$ and electrochemical
potentials $\mu_r=\mu \pm V/2$, where $V$ is the applied bias voltage.
The average chemical potential is defined such that $\mu=0$ corresponds to the
charge degeneracy point (including the change in vibrational
zero-point energy of the molecule).
The gate voltage then effectively varies $\mu$ relative to the
vibrational transition energies of the molecule.
The Hamiltonians $H_N, N=0,1$ describe the vibrational motion in the
neutral and charged electronic state, respectively, which gives rise
to the excitation spectra. This is depicted in Fig.~\ref{fig:pot}.
Here $x$ and $p$ ($[x,p]=i$) are the vibrational coordinate and
momentum, respectively, and $M$ is the effective mass.
Typically, in each charge state the lowest order approximation to
effective nuclear potential around its minimum is a harmonic
potential.
Importantly, both the equilibrium position and {\em frequency} depend
on the charge state, $\omega_0 \neq \omega_1$ and $\delta x \neq
0$. Which effect will be more important depends on microscopic details
of the molecule.
Since the ZPMs are different it is not obvious how to
define a single normalized coordinate.
It is convenient to introduce the geometric-mean frequency
$\Omega=\sqrt{\omega_0 \omega_1}$ as the vibrational energy scale.
We relate the shift to the dimensionless  parameter
$\lambda= \sqrt{M\Omega/2}\delta x$
and the frequency distortion to
$\alpha=\sqrt{\omega_0 / \omega_1}$
(i.e. $\omega_{0,1}=\alpha^{\pm}\Omega$).
Using the dimensionless coordinate $Q=\sqrt{M\Omega}x$ normalized to
the ZPM associated with $\Omega$
 and the conjugate momentum $P$, $[Q,P]=i$ we rewrite $H_N$ in the
form (see also \ref{app:fc})
\begin{equation}
  H_N         = 
     \frac{\omega_N}{2}
     \left[
       \frac{P^2}{\alpha^{\pm}} + \alpha^{\pm}\left( Q \pm \lambda / \sqrt{2} \right)^2
     \right]
     .
  \label{eq:HN}
\end{equation}
with $\pm$ for $N=0,1$.
We label the molecular eigenstates by $N_m$ and write their wave functions
as $\ket{N}\ket{m}_N$,  where
$\ket{m}_N$ is the vibrational eigenstate with $m=0,1,2,...$ quanta
excited on the potential surface for electronic state $\ket{N},N=0,1$:
$H_N \ket{m}_N=\omega_N(m+1/2)\ket{m}_N$.
Different regimes are characterized by comparing the change in the
classical elastic energies involved in the vertical transitions
 to each potential (see Fig.~\ref{fig:pot}) with the vibrational frequency
 of the final charge state.
(Equivalently one compares the ZPM of each potential with the relative
shift of the two potentials.)
The relevant dimensionless couplings are thus
 $\lambda^2\alpha \gtrless 1$ and
 $\lambda^2/\alpha \gtrless 1$.
Due to the spatial inversion symmetry of each potential about its minimum
the transport problem is invariant under $\lambda \rightarrow - \lambda$
and $(\alpha, \mu) \rightarrow  (\alpha^{-1}, -\mu)$ (see \ref{app:fc}).
Therefore, assuming $\lambda  \ge 0, \alpha \ge 1$
 only three regimes need to be considered
  (i)~$\lambda^2 < 1/\alpha$,
 (ii)~$1/\alpha < \lambda^2 < \alpha $, and
(iii)~$\alpha < \lambda^2$.
\\
{\em Transport.}
The tunneling of electrons with an excess energy provided by the bias voltage
drives the molecule out of electronic and vibrational equilibrium.
In the weak tunneling regime, i.e. $\Gamma \ll T$,
the occupation probabilities $P^N_m$ for $N$ electrons on the molecule and 
$m$ vibrational quanta excited may be described by a stationary master equation.
Neglecting for now relaxation of the vibrational states by the environment
(see Sect.~\ref{sec:relax}) we have:
\begin{eqnarray}
 & \dot{P}_m^0   =0&=\sum_{rm'}\left(
       W_{0m\leftarrow 1m'}^{r}P^1_{m'}-W_{1m'\leftarrow 0m}^{r}P^0_m
      \right)
     ,
     \nonumber \\
 & \dot{P}_{m'}^{1} =0&=        \sum_{r m}\left(
     W_{1m'\leftarrow 0m}^{r}P^0_{m}- W_{0m\leftarrow 1m'}^{r}P^1_{m'}
      \right)
    ,
\label{eq:dotP}
 \end{eqnarray}
 with transition rates due to tunneling to/from electrode $r=L,R$
 \begin{eqnarray}
 W_{1m' \leftarrow 0m}^{r} & =& \Gamma^{r} F_{m'm}    f_{r}( \mu_{m'm} )
 ,
 \nonumber \\
 W_{0m \leftarrow 1m'}^{r} &= & \Gamma^{r} F_{m'm}[ 1-f_{r}( \mu_{m'm} ) ]
 .
 \label{eq:rate}
 \end{eqnarray}
 where $f_r(E) \equiv (e^{(E-\mu_r)/T}+1)^{-1}$.
 The addition energies for the transition $1_{m'}\leftarrow 0_m$ are
 \begin{equation}
   \mu_{m'm}  = \omega_1 m'-\omega_0 m
              = \Omega (\alpha^{-1}m'-\alpha m )
   .
 \end{equation}
 and the Frack-Condon factors
 \begin{equation}
   F_{m'm}  = |_1\braket{m'}{m}_0|^2
   .
   \label{eq:fc}
 \end{equation}
 Where possible we will reserve $m$ and $m'$ for vibrational numbers
 of the charge state $N=0$ and $N=1$, respectively.
 The stationary current flowing out of reservoir $r=L,R$ is given by
 \begin{equation}
   I_r=\sum_{m m'} \left(
     W_{1m'\leftarrow 0m}^{r}P_{m}^0
     -W_{0m\leftarrow 1m'}^{r}P_{m'}^1
   \right)
   \label{eq:I}
 \end{equation}
The probabilities are normalized, $\sum_{N m}P^N_m=1$,
 and the current is conserved, $I_L+I_R=0$.
\\
{\em Spectral features.}
Due to the $\mu$ and $V$ dependence of the transition rates~(\ref{eq:rate})
the current will change whenever a resonance $\mu_{r}=\mu_{m'm}$  is crossed.
For $r=L,R$ this defines a line with negative/positive slope in $(\mu,V)$ plane
for $V>0$ (which we consider from hereon)
where a new electron tunneling process becomes possible.
Here the molecule can change its {\em vibrational} energy by an amount
 $\mu_{m'm}$.
Without distortion ($\alpha=1$) this resonance condition only depends
on the change in vibrational number $m'-m$.  For instance, once the
transition $0_0 \rightarrow 1_{k}$ is energetically allowed for
some fixed integer $k$, transitions $0_m \rightarrow 1_{m +k}$ are allowed
for {\em all} $m=1,2, \ldots$.  This infinity of processes becomes
energetically allowed at a single resonance line.  In combination with
other allowed transitions a {\em cascade} of transitions gives access
to arbitrarily high excited states~\cite{Nowack05}
and results in a divergence of the average phonon-number when we let
$\lambda \rightarrow 0$~\cite{Koch05a} for fixed applied voltages.
For $\alpha \neq 1$ the {\em mean} of the two vibrational numbers
also enters into the resonance condition since
$\mu_{m'm}/ \Omega =
   (\alpha+\alpha^{-1})(m'-m )/2
 - (\alpha-\alpha^{-1})(m +m')/2$
 The cascades of transitions are now switched on in a sequence of steps.
\\
{\em Intensities.}~The sign and intensity of the change in the current
at the resonances is determined by the rates $\Gamma^{r}$ for
tunneling to/from electrode $r=L,R$ 
and the Franck-Condon (FC) factors $F_{m'm}$.
The FC-factors take into account that the nuclear potential is
altered when the molecule becomes charged.
Their energy dependence through the vibrational numbers $m,m'$
is typically dominant over that of the
rates which we take to be constants $\Gamma^{r} = 2\pi  |t_r|^2 \rho_r$
with density of states $\rho_r$ in electrode $r=L,R$.
In contrast to most transport models,
here the FC-factors are non-symmetric $F_{m'm} \ne F_{mm'}$.
In general this is the case when the nuclear potentials in the two charge states
are not identical up to a shift.
The sum rules $\sum_m F_{m'm}=\sum_{m'} F_{m'm}=1$ are due to the
completeness of each vibrational basis set $\{\ket{m}_0\}$ and
$\{ \ket{m'}_1 \}$.
These guarantee that the current and the total occupation
$P^N=\sum_m P^N_m$ of each charge state $N$ will saturate at
large bias voltage to the electronic limit (i.e. without the vibration)
$I_{e}=(\sum_r 1/\Gamma^r)^{-1}$ and
$P^{0,1}_{e}=\Gamma^{L,R}/(\sum_r 1/\Gamma^r)$,
provided the FC-factors are bias voltage independent (cf.~\cite{McCarthy03}).
\subsection{Franck-Condon factors - Classical and quantum features
\label{sec:fc}
}
 The FC-factors for any $\lambda$ and $\alpha$ can be calculated
 analytically by disentangling the unitary transformation
 which maps the oscillator $H_1$ onto the oscillator $H_0$.
 The expressions and their derivation are deferred to
 \ref{app:fc} and we will focus here on their essential features.
 \begin{figure}
   \begin{center}
     \includegraphics[scale=0.95]{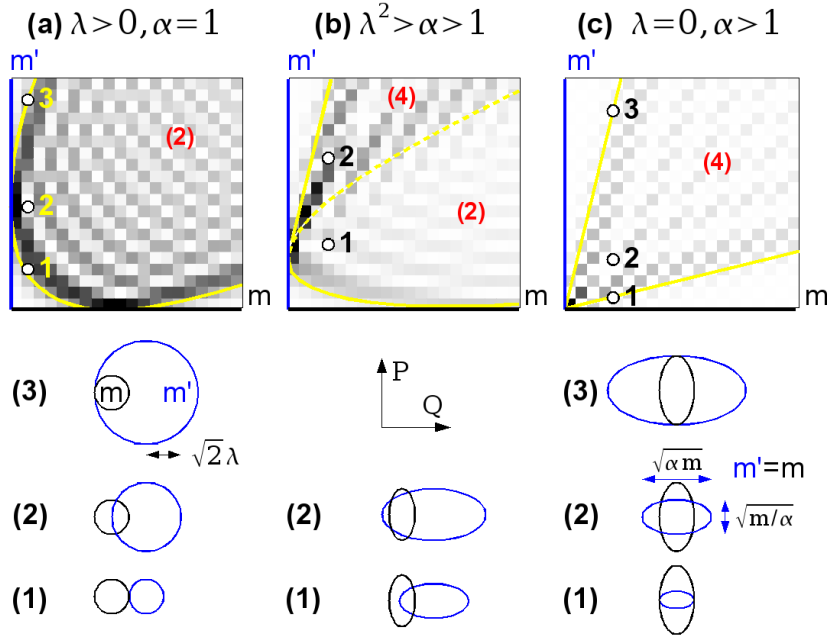}
     \caption{ \label{fig:fc}
       Franck-Condon factors in grayscale (black = 1,white=0).
       The quasi-classical boundary curves (yellow lines) separate
       classically allowed and forbidden transitions (see \ref{app:class}).
       Sketched below each panel are the classical phase space $(Q,P)$
       trajectories of the vibration in charge state $N=0$ (black) and
       $N=1$ (blue)
       with energy $m\omega_0$ and $m'\omega_1$.
       The red numbers $(2)$ and $(4)$ in the grayscale plots indicate the
       number of intersections of these trajectories.
       (a) $\alpha=1   , \lambda=3.0$.
       For strong shift, $\lambda^2 \gg \alpha$ the FC-factors are
       suppressed for small $m',m$ and exponentially increase as the
       classically allowed region is approached.
       (b) $\alpha=2.05, \lambda=3.0$.
       For combined shift and distortion the FC-factors are asymmetric
       in $m,m'$ (although they are symmetric in the limiting
       cases (a) and (c) due to special symmetries).
       (c) $\alpha=2.05, \lambda=0$.
       For strong distortion, $\lambda^2 \ll 1/\alpha$ the FC-factors rapidly
       decrease with $m,m'$ and in addition oscillate due to the
       parity selection rule (checkerboard pattern).
     }
   \end{center}
 \end{figure}
In Fig.~\ref{fig:fc} we plot the FC-factors in gray-scale for three
representative cases.
Their large-scale dependence on the vibrational numbers $m,m'$ follows
from quasi-classical arguments as is discussed in detail in
\ref{app:class}.
In the case of only a large shift ($\lambda^2 > 1, \alpha=1$, 
Fig.~\ref{fig:fc}(a)), the FC-factors are basically nonzero only
in a classically allowed region bounded by the so-called Condon
parabola~\cite{Herzberg,Nowack05}.
The maximal values occurs at $m=\lambda^2,m'=0$ and
$m=0,m'=\lambda^2$.
For $\lambda \rightarrow 0$ this parabola narrows down to a line,
$F_{m'm} \rightarrow \delta_{m'm}$.
For only a large distortion ($\alpha \gg 1, \lambda=0$,
Fig.~\ref{fig:fc}(c)), the classically allowed region has two linear
boundaries (which coincide for $\alpha \rightarrow 1$) and the maximal
value occurs for $m=m'=0$.
When both a shift and a distortion occur (Fig.~\ref{fig:fc}(b)) the
boundary curve is partially linear and partially parabolic.
Additionally, two classically allowed regions of different intensity
can be distinguished, corresponding to a difference in possible
classical motions.
The global maximum occurs for $m=0,m' = \lambda^2/\alpha$ and a local maximum
for $m=\lambda^2 \alpha,m'=0$,
corresponding to the two {\em different} classical elastic energy
scales.
It is important to note that for $\alpha > 1$ one can have a shift
which is large relative to the ZPM of one potential
 ($\lambda^2 \alpha > 1$) but still small relative to the ZPM of the other
 ($\lambda^2 / \alpha > 1$).
In this case interference effects may occur in the FC-factors which
are directly observable in the electronic transport (see
Sect.~\ref{sec:asym}).
\\
The distortion breaks the symmetry between the neutral and
charged state in two respects:
 the vibrational excitation spectra of the neutral and charged
 state become asymmetric
and the FC-factors become asymmetric as function of the
{\em energies} $E_{m}  = m \omega_0
               ,E_{m'} = m'\omega_1$
(not shown). Note that this is also the case for $\lambda=0,\alpha>1$, even
though in this case $F_{m'm} = F_{m m'}$ (see Fig.~\ref{fig:fc}(c)).
Together these lead to a lack of inversion symmetry of the current
with respect to the gate energy $\mu \rightarrow - \mu$ even in the
case $\Gamma^L=\Gamma^R$.
\\
The qualitative classical understanding of the FC-factors allows one
to understand the dependence of the occupations and the current on the
applied voltages in  detail using figures like Fig.~\ref{fig:fc}.
This is discussed in~\ref{app:vibdist}.
For instance, for strong shifts ($\lambda^2 \gg 1,\alpha=1$) one can
explain the blockade of the current at low voltage~\cite{Koch04b},
negative (NDC) instead of positive differential conductance (PDC), and
even sharp current peaks~\cite{Nowack05} in terms of a feedback
mechanism in the vibration-assisted transitions between the molecular
states.
The detailed variations of the FC-factors within the classically allowed
region are not needed for this.
Although the current steps at discrete energies are due to the
quantized nuclear motion the variation of their sign {\em and}
intensity follow from quasi-classical features of the FC-factors.
However, when a distortion is present {\em interference} effects in the 
sign and  intensity of the conductance resonances may appear.
For instance, for purely distorted potentials ($\lambda=0$) the spatial
inversion symmetry of the vibrational wave function cannot be changed
by the electron tunneling.
This leads to a strict parity-selection rule for the FC-factors (see
\ref{app:fc}):
\begin{equation}
  F^{\lambda=0}_{m'm}=0~{\rm unless}~m'-m={\rm even}
  \label{eq:parsel}
\end{equation}
This is visible as a checkerboard pattern in Fig.~\ref{fig:fc}(c).
For weakly shifted but distorted potentials the tunneling rates
can still vary significantly when the vibrational number changes by
only {\em one} (see Sect.~\ref{sec:parity}).
This leads to even-odd effects in the intensity and sign of the conductance.
For a intermediate shift of the potentials the decay rate of a {\em
single} low-lying excited state can be coherently suppressed while its
rate of population is coherently enhanced.
This is possible only when the nuclear ZPM of the two potentials are
sufficiently different.
As a result the transport is suppressed in a large regime of applied voltages.
Both effects require asymmetric excitation spectra i.e. a distortion.
This indicates why interference effects in the FC-factors played no role in
 previous works.
\subsection{Relaxation
  \label{sec:relax}
}
The FC-factors strongly influence the type of non-equilibrium vibrational
distribution which the transport current induces on the molecule~\footnotemark.
\footnotetext{
  We point out that for $\lambda=0$ the model makes little sense
  physically without relaxation
  since for $\alpha\ne 1$ the even and odd $m$ states can not be mixed
  by electron tunneling processes
  and for $\alpha=1$ the vibrational number cannot change since
  $F_{m'm}=\delta_{m'm}$.
  The master equation~(\ref{eq:dotP}) in these case has no unique
  stationary solution (i.e. the solution depends on initial conditions).
  This artifact immediately disappears when introducing a finite
  $\lambda$ or relaxation.
}
To be able to identify such effects qualitatively we compare our results with those
where vibrational excitations on the molecule can relax by
coupling to an environment of oscillators.
For simplicity we assume that the spectral function of the environment
is a constant $\gamma$ and the temperature is equal to that of the
electrodes, $T$.
For weak coupling to this environment, $\gamma \ll {\rm min} \left\{
\omega_N \right\}$ , its influence can be included through an
additional term
 $\sum_n
 (
    W_{Nm \leftarrow Nn}P^N_n
  - W_{Nn \leftarrow Nm}P^N_m
 )$
to the right-hand side of equation (\ref{eq:dotP}) for $\dot{P}^N_{m}$
without altering the expression for the current~(\ref{eq:I}).  The
rates are given by
\begin{equation}
  W_{Nn \leftarrow Nm}  =  \gamma [\pm b(\omega_N (n-m) )]
\end{equation}
for $ n \gtrless m$ and $N=0,1$ where $b(E)=(e^{
E/T}-1)^{-1}=-(1+b(-E))$.  The relaxation can be either
strong ($\gamma > \Gamma$) or
  weak ($\gamma < \Gamma$) relative to the tunneling as long as both
are smaller than $T$ and ${\rm min} \left\{  \omega_N \right\}$.
What is of interest here is that non-equilibrium effects of different
physical origin disappear at different characteristic strengths of the
relaxation $\gamma \lesssim \Gamma$
i.e. they have a specific sensitivity to relaxation processes.
The case of strong relaxation will not be discussed except for the
important fact that in this limit NDC effects vanish
in any single orbital model regardless of the
FC-factors~\cite{Nowack05}.
This may be shown  using an equilibrium ansatz for the
vibrational distribution~\cite{Braig03a,Mitra04b}.
Thus in limit of weak tunneling and weak coupling to the environment
considered here {\em NDC implies non-equilibrium}.
\section{Results
\label{sec:results}
}
The stationary current $I$
 (Eq.~(\ref{eq:I})) and differential conductance $dI/dV$ are presented
 for symmetric tunneling rates $\Gamma^{r}=\Gamma,r=L,R$ and temperature
 $T/ \Omega=0.01$.
 Gray-scale plots of $dI/dV$ have different linear scale
 factors for $dI/dV \gtrless 0$ to clarify voltage conditions for which
 NDC effects occur.
 Their magnitudes can be appreciated from the presented $I(V)$ curves
 or from the text.
 We will consider parameter values which are well separated to keep
 the discussion simple.
 At this point a general conclusion can already be made:
 for all values of $\lambda$ for which we present results no NDC is
 visible if we set $\alpha=1$.
 The occurrence of NDC in all cases where $\alpha \ne 1$ indicates
that a {\em distortion} enhances {\em non-equilibrium} vibrational
effects, however via several different mechanism which we will now analyze.
\subsection{Nearly symmetric excitation spectra:  $\alpha^2 < 2$
\label{sec:sym}}
For moderate values of $\alpha^2 < 2$ only one vibrational
excitation of $N=1$ lies below the first one for $N=0$ (cf. Fig.~\ref{fig:pot}).
The low energy spectra in the two charge states may thus be
characterized as nearly identical.
Interestingly, due to the slight asymmetry the current provides direct
information on the changes in vibrational distribution which remain
hidden without a distortion.
Some effects of interference in the FC-factors may be identified.
\subsubsection{{Small shift} $\lambda^2 \ll 1/\alpha,\alpha$:
  Broadening of the vibrational distribution
\label{sec:casc}}
\begin{figure}
  \begin{center}
    \includegraphics[scale=0.7]{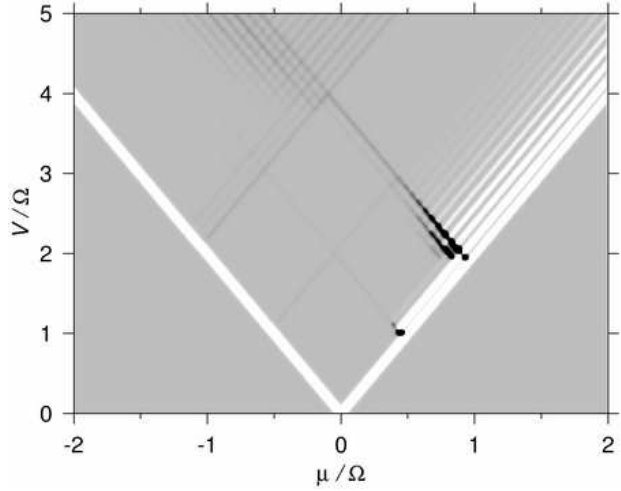}
    \caption{
      \label{fig:G_casc}
      Differential conductance for $\alpha=1.05, \lambda=0.01$.
      White and black relative to the grey background indicate
      positive and negative intensity of $dI/dV$. The same holds for all
      subsequent grayscale plots.
    }
  \end{center}
\end{figure}

\begin{figure}
   \begin{center}
     \includegraphics[scale=0.3,angle=-90]{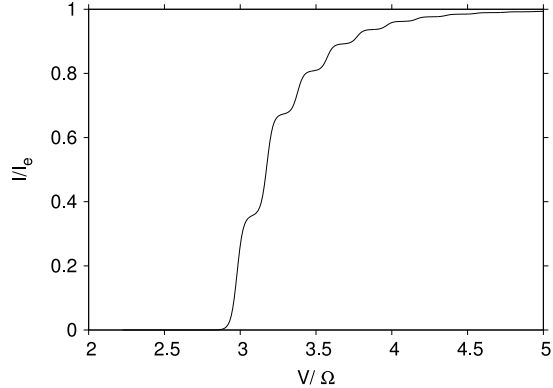}
     \caption{
       \label{fig:I_casc}
       $I(V)$ normalized to the electronic limit $I_e$ for
       $\mu/\Omega=1.5$ in Fig.~\ref{fig:G_casc}. The current steps are
       spaced by the frequency {\em difference} $\omega_0-\omega_1$.
     }
   \end{center}
 \end{figure}

\begin{figure}
  \begin{center}
    \includegraphics[scale=0.7]{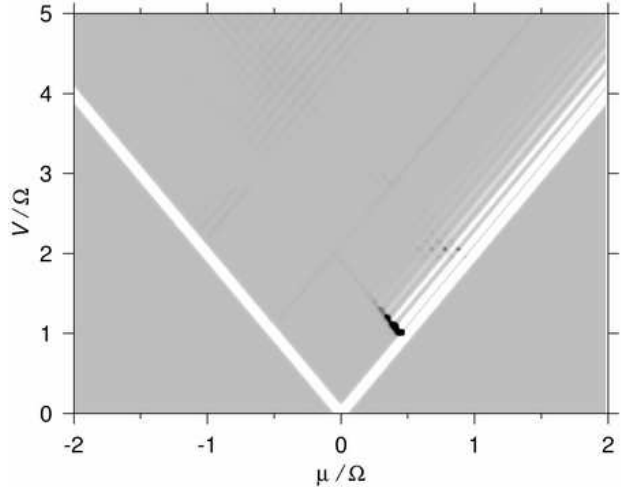}
    \caption{
      \label{fig:G_casc2}
      Differential conductance for $\alpha=1.05, \lambda=0.1$.
    }
  \end{center}
\end{figure}

For $\alpha=1$ the $dI/dV$ is completely featureless apart from the
ground-state transition lines.
A distortion causes the ground-state resonance line to split in many
excitation lines which can be resolved at low temperature, as can be
seen in Fig.~\ref{fig:G_casc}.
The current, shown in Fig.~\ref{fig:I_casc},
is strongly modulated on the new small energy scale
$\omega_{0}-\omega_{1}$ by which the resonances are separated.
This modulation is expected since the tunneling rates strongly
decrease with increasing energy.
The featureless result for $\alpha=1$ is rather special since it is
due to the exact symmetry of the excitation spectra for $N=0$ and
$N=1$.
\begin{figure}
  \begin{center}
    \includegraphics[scale=0.5,angle=-90]{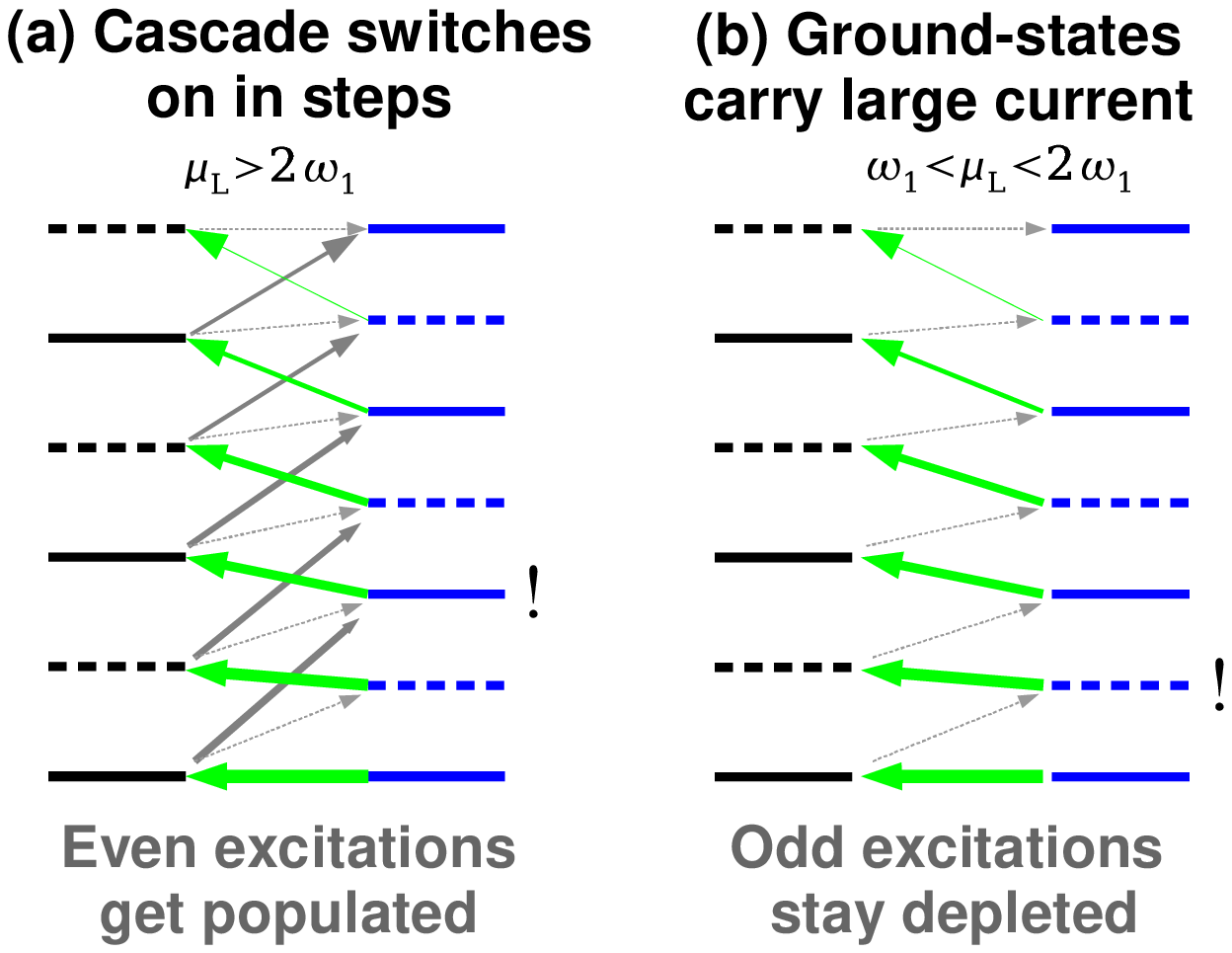}
    \caption{
      \label{fig:scheme-casc}
      Transitions relevant to the right in
      Fig.~\ref{fig:G_casc}.
      In all this kind below
      black/blue levels correspond to ground- and
      excited vibrational states of the neutral/charged molecule,
      $0_m,m=0,1,2,\ldots$/ $1_{m'},m'=0,1,2,\ldots$.
      The parity of the states is indicated by the line style (full=even,
      dashed=odd).
      Arrow thickness indicates the relative strength of the FC-factors.
      Parity forbidden/allowed transitions are indicated by a
      full/dashed arrow.
      Not all transitions are indicated.
      (a) For $\mu_L>2 \omega_1$  all grey transitions are energetically
      allowed and only the missing links (green) in the cascade need to be
      switched on, one by one starting from below, at the resonance lines
      $-\mu_R=k(\omega_0-\omega_1)$ for $k=1,2,\ldots,[1/(\alpha^2-1)]+1$.
      After the last transition is possible the scheme repeats for the
      higher lying states i.e. all states are energetically accessible and a
      broad vibrational distribution is established.
      (b) For $2\omega_1> \mu_L>\omega_1$ the progression does not
      occur due to the parity effect.
    }
  \end{center}
\end{figure}
The progression of lines maps out the stepwise lengthening of cascades
of transitions (Fig.~\ref{fig:scheme-casc}(a)).
Once $[1/(\alpha^2-1)]+1$ ($[x]=$ integer remainder of $x$)
of such resonances  have been traversed
(more than $2$ for $\alpha < \sqrt{2}$)
, the cascade is infinitely long i.e. any state can be reached.
This is the case for $-\mu_R > \omega_0$.
The broadening of the vibrational distribution is thus sharply
controlled by the bias voltage in this region.
The individual processes which lead to the broad non-equilibrium
vibrational distribution for $\lambda \ll 1$,
discussed in~\cite{Koch05a}, may thus
be identified in the transport current if the vibrational excitation
spectrum is charge dependent due to a distortion.
From Fig.~\ref{fig:fc}(c) one sees that for finite
$\alpha$ the FC-factors do not collapse onto a line
$F_{m'm}=\delta_{m'm}$ as we let $\lambda \rightarrow 0$.
We therefore argue that the scaling of the vibrational distribution
width found for $\alpha=1$ in~\cite{Koch05a} may break down for
$\alpha \ne 1$ below some cut-off value for $\lambda$ (which depends
on $\alpha$).
\\
For $2\omega_1 > \mu_L > \omega_1$ the progression seems to
break down after the first large step.
This is due to the suppression of all rates between even and
odd excitations cf. Eq.~(\ref{eq:parsel}), as
Fig.~\ref{fig:scheme-casc}(b) illustrates.
Except for one sharp feature and a weak NDC effect
the resonance lines corresponding to
quasi-forbidden transitions are missing.
Pronounced NDC occurs when the region
$\mu_L             > \omega_1
 ,\omega_0-\omega_1 > -\mu_R$ is reached:
the change in the current is $1/3$ of its maximal value.
The occupation of state $1_1$ via a {\em quasi-forbidden} transition
causes the drop in current since the probability is redistributed over the
three states $0_0,1_0$ and $1_1$ and the latter does not contribute to the
current since $F_{10} \ll F_{00}$.
Once the transition $0_1 \leftarrow 1_1$ is energetically allowed all
excited states are de-populated due to the quasi-selection rule for the
FC-factors:
$F_{10} \ll F_{11}$.
The current nearly reaches its maximal value $\Gamma/2$.
For $\lambda^2 \lesssim 1/\alpha$ the selection rule is sufficiently
weakened that the progression already starts
at $\mu_L=\omega_1$ as seen in Fig.~\ref{fig:G_casc2}.
\\
The progression is a non-equilibrium effect as it consists only of
transitions between excited states. The number of visible lines is
indicative of the average vibrational number and is reduced
when a relaxation rate $\gamma > 0$ is switched on (not shown).
We point out that in Fig.~\ref{fig:G_casc} the energy separation in this
progression of excitations could be mistaken for a vibrational
frequency.
Also one might infer erroneously a coupling $\lambda \lesssim 1$
since many resonances with decreasing intensity can be resolved.
In fact the extent of the progression shows an 
dependence on the shift opposite to that of a usual FC-progression:
it becomes shorter with increasing $\lambda$.
Crucial for a correct identification are furthermore the featureless
low bias region ($|\mu_r|< \omega_0,\omega_1$) and the NDC effects.
\subsubsection{{Large shift} $1/\alpha,\alpha \ll \lambda^2$:
 Trapping in the vibrational ground state
\label{sec:trap1}
}
\begin{figure}
  \begin{center}
    \includegraphics[scale=0.7]{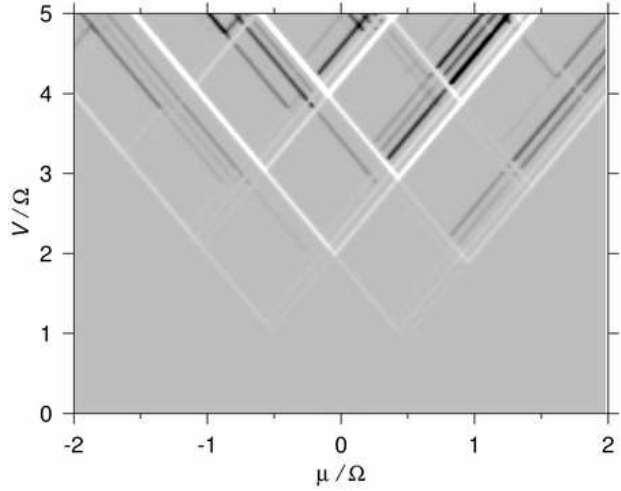}
    \caption{
      \label{fig:G_trap1}
      Differential conductance for $\alpha=1.05, \lambda=3.0$.
    }
  \end{center}
\end{figure}

For strong shift the parity of the nuclear wave functions
plays no role:
in Fig.~\ref{fig:G_trap1} 
resonance lines $\mu_L  = m \omega_1$
            and $-\mu_R = m \omega_1$ for $m=$~odd are now
clearly visible.
\begin{figure}
  \begin{center}
    \includegraphics[scale=0.3,angle=-90]{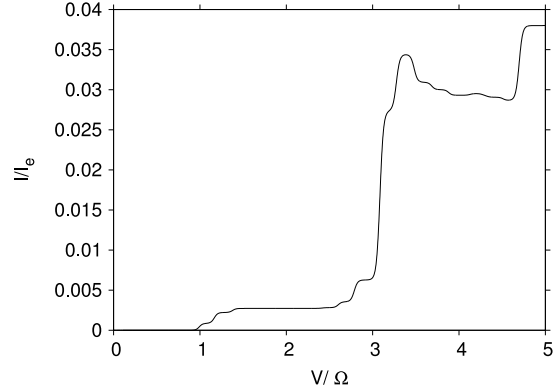}
    \caption{ \label{fig:I_trap1}
       $I(V)$ normalized to the electronic limit $I_e$
      for $\mu/\Omega=0.5$ in Fig.~\ref{fig:G_trap1}.
      We note that the average vibrational number shows the same
      marked peak as function of the bias voltage.
    }
  \end{center}
\end{figure}
\begin{figure}
  \begin{center}
    \includegraphics[scale=0.6,angle=-90]{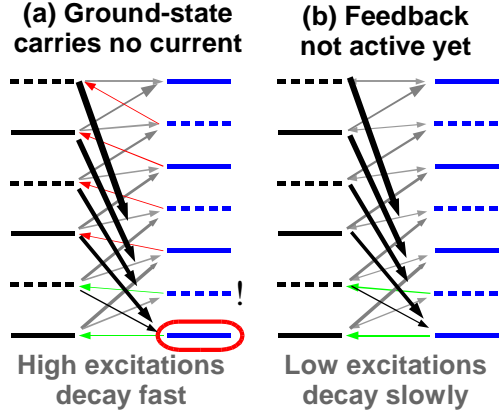}
    \caption{
      \label{fig:scheme-trap1}
      Transitions relevant to the right in
      Fig.~\ref{fig:G_trap1}.
      As in Fig.~\ref{fig:scheme-casc} the missing links (green and
      red) in the  cascade are switched on.
      However, in this case the molecule becomes trapped in the ground
      states (not contributing to the current) for most applied voltages:
      once excited states become slightly occupied they feed back
      strongly into the ground state (black arrows) since
      the transition rates exponentially {\em increase}
      with the difference in vibrational energy (opposite to
      Fig.~\ref{fig:scheme-casc}), see Fig.~\ref{fig:fc}.
      Due to the gate asymmetry, $\mu > 0$, the ground state can only
      decay via processes with much smaller rates.
      Due to the distortion of the spectrum this feedback mechanism is
      not yet active at the first two resonances (green arrows), causing the
      current to ``overshoot'' resulting in the marked peak in
      Fig.~\ref{fig:I_trap1}.
    }
  \end{center}
\end{figure}
For $\alpha=1$ a strong shift $\lambda^2 \gg 1$ of the potentials is
well-known to suppress the ground state transition
line~\cite{Flensberg03,Mitra04b,Koch04b} and redistribute the weight
into a FC-progression of conductance resonances.
At this point it should be emphasized that the suppression is a
non-equilibrium effect even though the vibrational distribution has
its main weight in the ground states.
This is evidenced by the disappearance of the NDC and current
peaks~\cite{Nowack05} which occur for values of $\lambda \gtrsim 4$
and asymmetric gate energy ($|\mu| > \omega$)
and the  enhancement of the suppression~\cite{Koch04b} upon full
vibrational equilibration.
\\
The main effect of the distortion comes from the asymmetric spectrum.
What is remarkable in Fig.~\ref{fig:G_trap1} compared to
Fig.~\ref{fig:G_casc} is that most of the resonances at higher bias
with spacing on the small scale $\omega_{0}-\omega_{1}$ lead to NDC,
whereas the excitations on the larger scale $\omega_{0},\omega_{1}$
all correspond to PDC.
The same cascade of transitions discussed in Sect.~\ref{sec:casc} now
stabilizes the lowest vibrational states~\cite{Nowack05} which
contribute little to the current.
However, due to the small mismatch of the excitation spectra 
the current can first reach a high value which is subsequently
reduced to the value it has for $\alpha=1$
when the cascades are switched on, see
Fig.~\ref{fig:I_trap1}.
This happens repeatedly with increasing bias.
The effect here should thus not be characterized as current
suppression.
Rather, the current is {\em enhanced} relative to the case  $\alpha=1$
where the feedback always dominates the current.
Due to the slightly asymmetric spectra
the trapping in the ground state is ``postponed''.
Markedly, the NDC lines have positive/negative slope for $\mu \gtrless 0$ which
is {\em opposite} to that of the NDC occurring for very large $\lambda$ and
$\alpha=1$~\cite{Koch04b,Nowack05}.
\subsection{Asymmetric spectra: $\alpha^2 > 2$
\label{sec:asym}}
For sufficient distortion, $\omega_0 > 2 \omega_1$, two or
more excitations of the $N=1$ charge state lie below the first
excitation in the $N=0$ state i.e. there is a true asymmetry between
the spectra of the two charge states.
This brings about a simplification: most of the features discussed
below are qualitatively reproduced when truncating the spectrum at
energies above the larger vibrational frequency $\omega_0$,
retaining only the states $m=0,1$ and $m'=0,\ldots,[\alpha^2]+1$
for $N=0$ and $N=1$ respectively.
Due to the presence of the $N=1$ low-lying states interference effects in the
FC-factors also gain importance.
For small shift the quasi-conservation of the nuclear wave function
parity suppresses the electron tunneling between {\em all}
even and odd vibrational states of $N=0$ and $N=1$.
For larger shifts $\lambda^2 \gtrsim 1/\alpha$
interference may suppress the decay rate of a
{\em single} low-lying excitation
 and  {\em simultaneously} enhance the rate at which this state is populated.
This concerted effect of constructive and destructive interference is
due to the opposite parity of the $N=0$ vibrational ground- and excited
state.
\subsubsection{{Small shift} $\lambda^2 \ll 1/\alpha$: Parity effect
\label{sec:parity}}
\begin{figure}
  \begin{center}
    \includegraphics[scale=0.7]{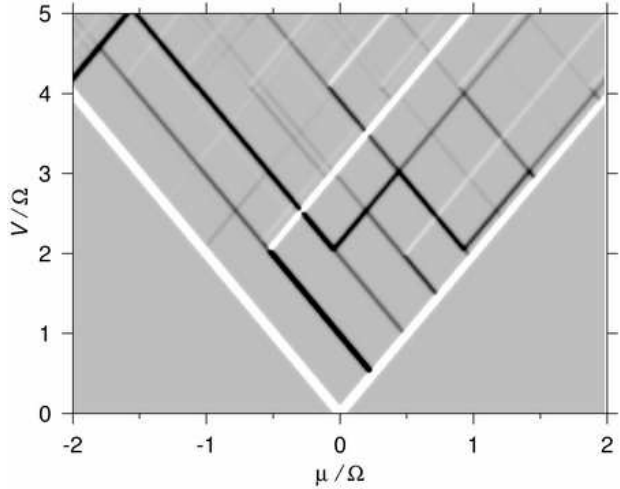}
    \caption{ \label{fig:G_par}
      Differential conductance for $\alpha=2.05, \lambda=0.01$.
    }
  \end{center}
\end{figure}
\begin{figure}
  \begin{center}
    \includegraphics[scale=0.3,angle=-90]{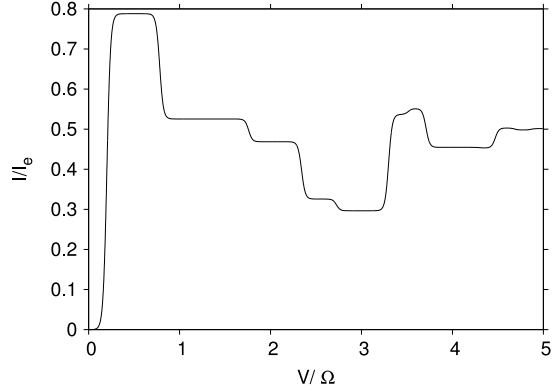}
    \caption{ \label{fig:I_par}
      $I(V)$ normalized to the electronic limit $I_e$
      for $\mu/\Omega=0.1$ in Fig.~\ref{fig:G_par}.
      The first three current plateaus values are describe by
      Eq.~(\ref{eq:Iplat}). In particular, for $\alpha^2 \gg 1$ the
      initial current drop by a factor 1/3 signals the redistribution
      of the probability from two to three states.
    }
  \end{center}
\end{figure}
\begin{figure}
  \begin{center}
    \includegraphics[scale=0.6,angle=-90]{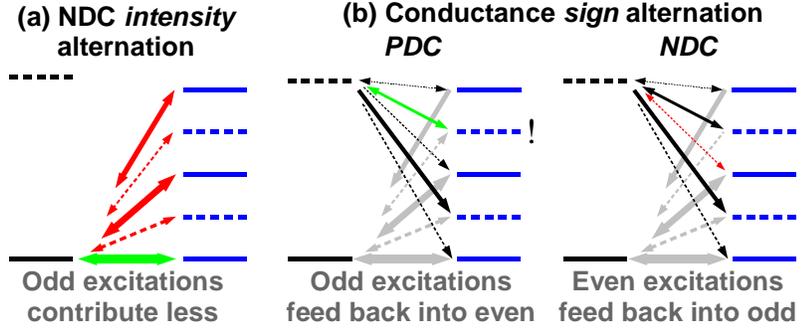}
    \caption{ \label{fig:scheme-par}
      Transitions relevant to the parity effects in Fig.~\ref{fig:G_par}.
      The closely spaced excitations belong to charged state $N=1$,
      the levels with large spacing (only two shown) belong to $N=0$.
      (a) Successive occupation of the non-contributing states
      $1_{m'}$ leads to NDC lines at low bias.
      The variation of the NDC intensity reflects wether the
      transition is parity forbidden (dashed line) or allowed (full line).
      (b) The successive opening of escape processes (green and red
      arrows) from the states not contributing to the current gives PDC for even
      and NDC for odd excitations.
    }
  \end{center}
\end{figure}
The $dI/dV$ in Fig.~\ref{fig:G_par} has two distinctive features.
Below the threshold voltage where $0_1$ can not yet be reached
(horizontal zig-zag line),
all resonance lines (with negative slope) correspond to NDC with an {\em
intensity} which alternates.
Above the threshold resonances  lines (with positive slope) appear with
alternating {\em sign} of the differential conductance.
Both effects derive from the parity quasi-selection rule
incorporated in the FC-factors and are systematic: as one increases
$\alpha^2$ by one, the additional resonances appearing below and above
the threshold voltage follow the above pattern.
We explain this parity effect using Fig.~\ref{fig:scheme-par}.
\\
{\em NDC intensity alternation.}
All the low voltage resonances correspond to NDC since the low-lying
excitations of $N=1$ become successively occupied
(Fig.~\ref{fig:scheme-par}(a))
without contributing significantly to the current
(the FC-factors $F_{m'0}$ rapidly decrease with $m'$).
Since the rates at which each excitation is populated and depleted are the
{\em same} (although small),
the ground state occupation is reduced $\propto 1/V$ resulting in a lower
current at higher bias.
However, the FC-factors $F_{m'0}$ also oscillate:
transitions from the ground state $0_0$
to odd-$m'$ states $1_{m'}$ are strongly
suppressed relative to those with even $m'$,
$F_{(2n+1)0} \ll F_{(2n)0}$ for $n=1,2,\ldots$.
Remarkably, the quasi-forbidden transitions  appear as
anti-resonances in the differential conductance (instead of missing
resonances) which modulate the current {\em stronger} than allowed
transitions. This effect is related to the non-equilibrium conditions
(see below).
This may be explicitly demonstrated from the expression
for current plateau $k=1,2,\ldots$ where $k \propto V$
\begin{equation}
  I_k =
  \left[
    (k+1)/\Gamma^R + 1/\Gamma^L
  \right]^{-1}
  \sum_{m'=0}^{k} F_{m'0}
  \label{eq:Iplat}
  .
\end{equation}
This result applies at low bias where states $m=0$ and states $m' \le
k$ are occupied and state $0_1$ is not yet reachable
(Fig.~\ref{fig:scheme-par}(a)).
For $\Gamma^{r}=\Gamma$ the current after an initial big step
decreases~$\propto 1/V$,
since the number of occupied excited levels which do not contribute to
the current grows $\propto V$.
Interestingly, for asymmetric coupling to the electrodes
 $\Gamma^L \ll \Gamma^R$
the depletion of the ground states is postponed:
the current initially increases in steps with $V$ to reach a maximum
around $V \propto \omega_1 (\Gamma^R / \Gamma^L) $ and then
decays $\propto 1/V$.  For negative bias the current shows no
such maximum.
\\
{\em NDC / PDC alternation.}
From the low bias resonances one can directly find the FC-factors
if the tunneling rates $\Gamma^{L,R}$ are known.
However, above the threshold voltage this is not the case anymore:
here multiple states from both charge sectors contribute in a
more complicated way (Fig.~\ref{fig:scheme-par}(b)).
Now escape from the non-contributing states $1_{m'}$ becomes possible.
Remarkably, this suppresses the current further at lines with positive
slope terminating at the NDC resonances due to quasi-{\em forbidden} transitions.
This is due to a {\em feedback} mechanism in the vibration assisted
transitions which effectively traps the system in the odd-parity
states  as explained in Fig.~\ref{fig:scheme-par}.
This is somewhat similar to the mechanism in the opposite case of
strong shifts (Sections~\ref{sec:trap1} and \ref{sec:trap2}) but
relies critically on the modulation of the rates due to the
quasi-conservation of parity.
\\
\begin{figure}
  \begin{center}
    \includegraphics[scale=0.7]{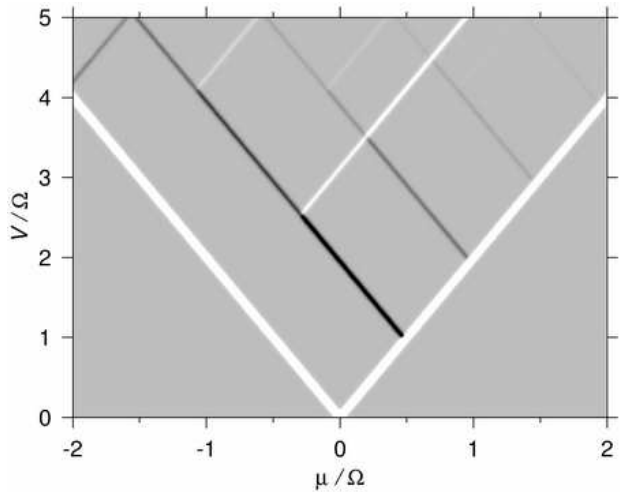}
    \caption{ \label{fig:G_par_relax}
      Differential conductance for $\alpha=2.05, \lambda=0.01$
      and finite relaxation rate $\gamma/\Gamma=0.1$.
      At this intermediate value of the relaxation only the odd
      NDC lines are suppressed, cf. Fig.~\ref{fig:G_par}.
      The spacing between the remaining even resonance lines with
      negative slope may be mistaken for the vibrational frequency of the
      excitations i.e. they mimic an effective frequency doubling.
      However, the resonance lines with positive slope are
      inconsistent with such an interpretation and hint at the missing
      resonances due to the parity effect.
    }
  \end{center}
\end{figure}
The allowed and forbidden excitation lines have different sensitivity
to relaxation processes and disappear in two stages.
When increasing the relaxation rate $\gamma$, in a first stage 
the strong NDC effects due to quasi-forbidden transitions
 become comparable in intensity with the NDC due to allowed ones
 and subsequently disappear as shown in Fig.~\ref{fig:G_par_relax}.
Thus similar to optical spectroscopy the parity selection rule now leads to
missing resonances in the spectrum.
The alternation of the sign of the resonances above the threshold voltage
also disappears since it is caused by asymmetries in the smallest rates.
Only in a second stage the remaining NDC lines due to the $m'=$even
states disappear.
\subsubsection{{Intermediate shift}~$1/\alpha < \lambda^2 < \alpha$:
  Coherent suppression
  \label{sec:cohsup}
}
\begin{figure}
  \begin{center}
    \includegraphics[scale=0.7]{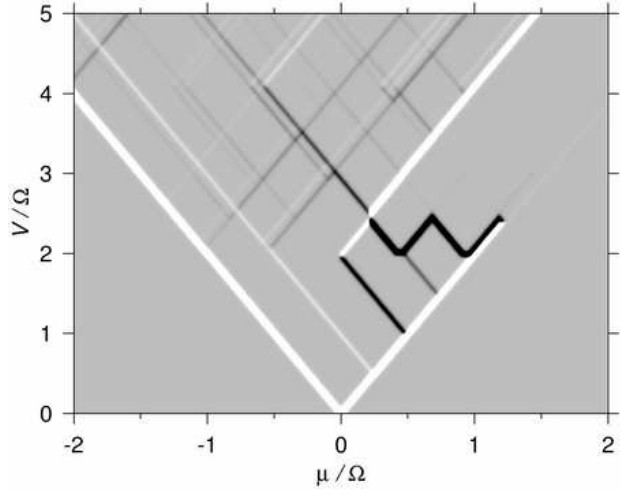}
    \caption{ \label{fig:G_cohsup}
      Differential conductance for $\alpha=2.05, \lambda=0.7$.
    }
  \end{center}
\end{figure}

\begin{figure}
  \begin{center}
    \includegraphics[scale=0.7,angle=-90]{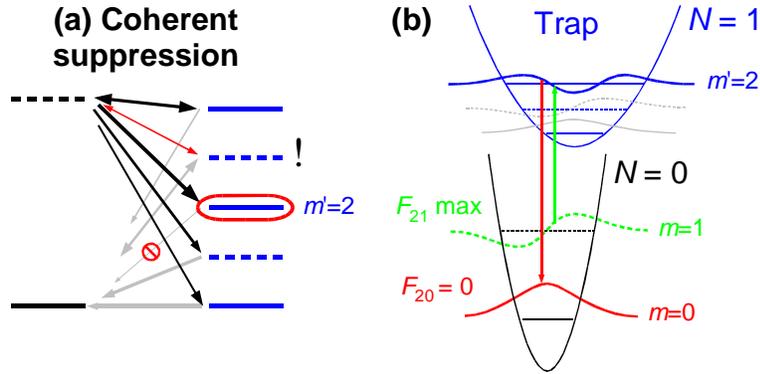}
    \caption{
      \label{fig:scheme-cohsup}
      (a) Transitions relevant to the 
      coherent suppression of the current on the right in
      Fig.~\ref{fig:G_cohsup}.
      The decay rate of excited state $1_2$ to the ground state $0_0$
      is suppressed due to destructive interference in the
      nuclear wave function overlap.
      State $1_2$ can thus not be reached directly,
      but due to the asymmetric spectra it can reached in
      three tunneling process via the excitations above it
      and the excited state $0_1$
      (for $V$ above the zig-zag threshold in Fig.~\ref{fig:G_cohsup}).
      The rate for the last process, $0_1 \rightarrow 1_2$ is
      maximized due to constructive interference.
      The coherently enhanced ratio of rates for going in and out of $1_2$
      suppresses the current.
      (b) Nuclear wavefunction overlap for $\lambda=\lambda_{(2)}$:
      the maximum and node of the states $0_0$ and $1_0$,
      respectively, align with the node of excitation $1_2$, thereby
      creating a low-energy blocking state.
    }
  \end{center}
\end{figure}

\begin{figure}
  \begin{center}
    \includegraphics[scale=0.7]{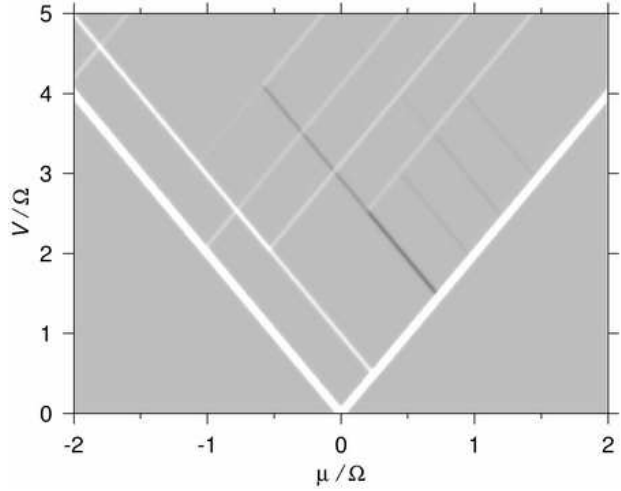}
    \caption{ \label{fig:cohsuprel}
      Differential conductance for $\alpha=2.05, \lambda=0.7$ and
      relaxation rate $\gamma/\Gamma=0.1$.
      Note that the excitation line of state with suppressed FC-factor
      is now missing and that the NDC at the zig-zag is weakened and moved to
      the straight line $\mu_L=3\omega_1$.
    }
  \end{center}
\end{figure}

Interestingly for any $\alpha>\sqrt{2}$ a drastic
suppression of the current occurs near special values of $\lambda$,
a prominent example being
\begin{equation}
  \label{eq:cohsup}
  \lambda_{(2)} = \frac{1}{2} \sqrt{ \alpha-\frac{1}{\alpha^3} }.
\end{equation}
As seen in Fig.~\ref{fig:G_cohsup} at large voltage bias the current is
completely suppressed beyond the ground-state transition line
$\mu_R=0$.
In this region the low-lying {\em excited} state $1_2$ becomes
completely occupied i.e. we have a bias driven population
inversion between vibrational states (cf.~\cite{Nowack05}).
The reason for this is twofold and is explained in
Fig.~\ref{fig:scheme-cohsup}(b):
due to {\em destructive} interference the rate of decay to the ground state
$0_0$ is suppressed (FC-factor $F_{20}=0$ for
$\lambda=\lambda_{(2)}(\alpha)$);
simultaneously {\em constructive} interference maximizes the tunneling rate
into this state from the excited state $0_1$.
This concerted effect is due to the opposite parity of the ground- and
excited state for $N=0$.
As soon as the excited state $0_1$ can be reached via some tunneling
processes the excited state $1_2$ becomes fully occupied and
suppresses the transport.
This happens at the  bias voltage threshold forming the horizontal
zig-zag line which we encountered above.
The transport is recovered only when direct
escape ($1_1 \leftarrow 0_2$) from the coherently blocked state
becomes energetically allowed i.e.
$-\mu_R > \omega_0 - 2 \omega_1$ (strong white line with positive above the
suppressed region in Fig.~\ref{fig:G_cohsup}).
The effect is {\em coherent} in the sense that both destructive and
constructive interference of the {\em nuclear} wave function are
responsible for the suppression of electron transport through the
molecule.
\begin{figure}
  \begin{center}
    \includegraphics[scale=0.82]{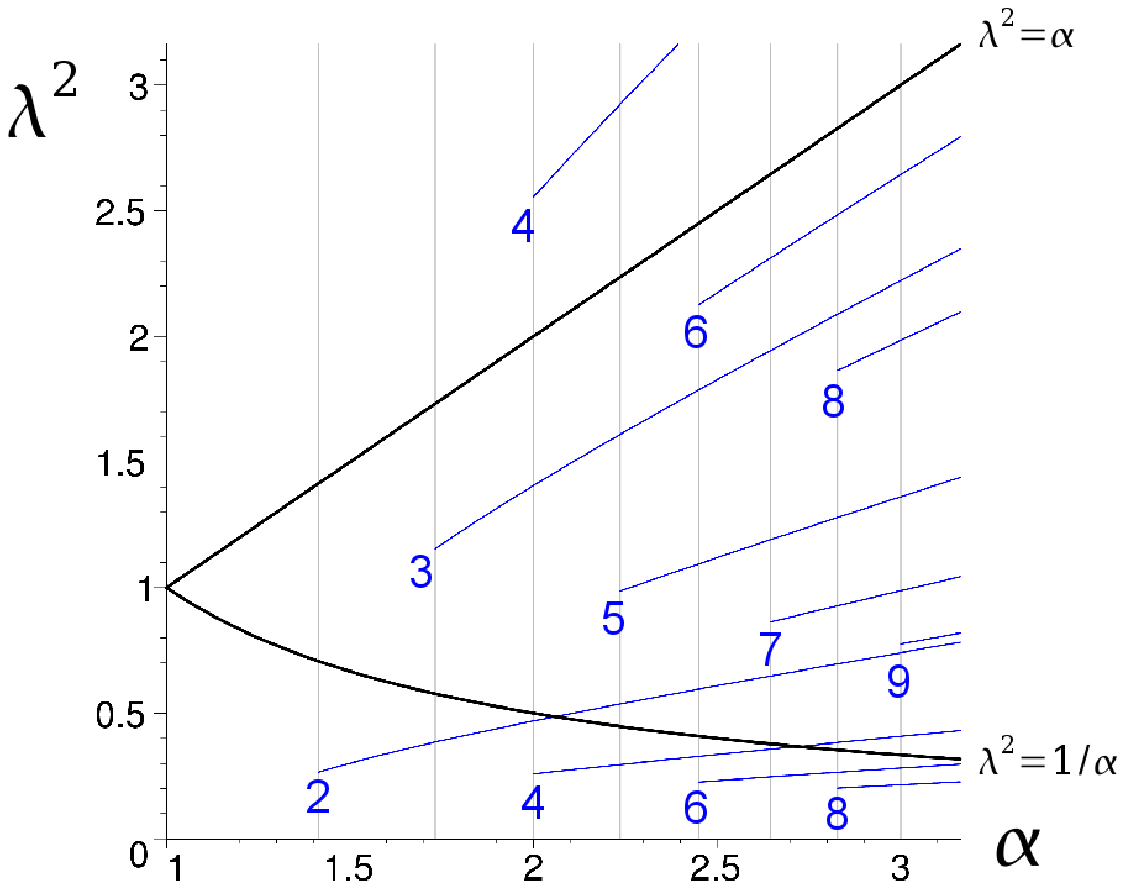}
    \caption{ \label{fig:zeros}
      Conditions for coherent suppression in the $(\alpha,\lambda^2)$
      plane.
      Regimes of
      weak ($\lambda^2 < 1/\alpha$),
      intermediate ($1/\alpha < \lambda^2 < \alpha$),
      and strong shift ($\alpha < \lambda^2$) are separated by black
      boundary lines.
      The blue curves labeled by integers $m'$ indicate the values of
      $\lambda$ for which the FC-factor $F_{m'0}$ vanishes for fixed
      $\alpha$.
      The curves are plotted only for those values of $\alpha$ for which
      state $m'$ lies below the first excitation of $N=0$ charge state
      and where the coherent suppression occurs.
    }
  \end{center}
\end{figure}
\\
In a similar way, the FC-factor $F_{m'0}$ of a higher excited state
 $1_{m'},m'=3,4,\cdots$ may vanish for some value of $\lambda$.
(For $m'=1$ this happens only for the trivial value $\lambda=0$.)
If in addition this state lies below the first excitation for $N=1$,
i.e. $\alpha^2 > m'$, this leads to a region of suppressed current
similar in shape to that in Fig.~\ref{fig:G_cohsup} but more narrow
(e.g. for $m'=3$ the width of the region is halved).
The lines in the $(\alpha,\lambda^2)$-plane where both these
conditions for the coherent suppression are satisfied are plotted in
Fig.~\ref{fig:zeros} for $m'=2-9$.
The curves are all of the form Eq.~(\ref{eq:cohsup}) but with a
prefactor which differs from $1/2$.
The region where this interference effect occurs is centered around
the regime $1/\alpha < \lambda^2 < \alpha$ and $\alpha > \sqrt{2}$,
where it is possible to have a shift which is larger than the
ZPM of the flattest potential but still smaller than that
of the steepest potential.
With increasing $\alpha$ the values of $\lambda$ where the coherent
suppression occurs start to abound and even proliferate into the
regime where the shift becomes strong, $\lambda^2 > \alpha$, see
Fig.~\ref{fig:zeros}.
With increasing $\alpha$ the number of such values rapidly increases
roughly $\propto \alpha^4$:
since state $1_{m'}$ has $m'$ nodes there are $[m'/2]$ zeros of
$F_{m'0}$ as a function of $\lambda$ for a given $\alpha$
and only the states $m'=2,\cdots,[\alpha^2]$ have energy $< \omega_0$.
Finally, we note that excited states with $m'>[\alpha^2]$ are not
expected to cause a coherent suppression effect since they always have
two (groups of) states with opposite parity to decay to.
It is highly unlikely that the decay rates to both types of states can
be suppressed simultaneously by a special choice of the shift
$\lambda$.
\\
Naturally the coherent effect is more sensitive to parameter
values than the quasi-classical trapping effect
(cf. Sect.~\ref{sec:trap1} and \ref{sec:trap2}).
This sensitivity has an interesting side to it.
(The effect of voltage dependence of $\lambda$ on the trapping effect
was considered in~\cite{McCarthy03}.)
Introducing only a {\em weak} dependence of the parameters, for
instance $\lambda$, on the bias voltage,
the current exhibits a pronounced dip down to
zero when tuning the parameters $\alpha$ and $\lambda$ with $V$
through the condition for coherent suppression~(\ref{eq:cohsup}).
In view of the many situations where this effect can occur
(Fig.~\ref{fig:zeros}) this is an interesting novel possibility to be
explored in single molecule devices.
\\
Similar to the parity effect, introducing relaxation affects the
transport in two stages. First the suppressed groundstate transition
line is restored and the forbidden transition line disappears as
depicted in Fig.~\ref{fig:cohsuprel}. An NDC effect related to the
state with the suppressed FC-factor is still present. In a second
stage this effect also vanishes.
Excited states $m'< [\alpha^2]$ have effective decay rate
$\approx (m'-1)\gamma$ and are therefore more sensitive to relaxation.
In summary: of all the excitations $1_{m'}$ the state $m'=2$ gives rise
to the strongest coherent suppression effect in the largest voltage region
and is the least sensitive to relaxation.
\subsubsection{{Large shift} $ \lambda^2 > \alpha$: 
Trapping in the vibrational ground state
\label{sec:trap2}}
\begin{figure}
  \begin{center}
    \includegraphics[scale=0.7]{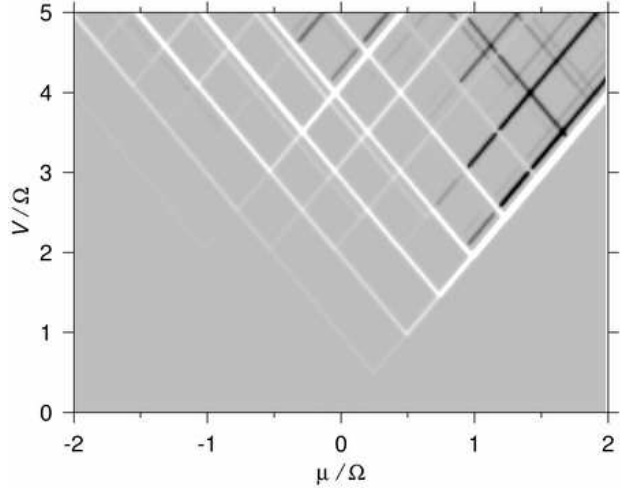}
    \caption{ \label{fig:G_trap2}
      Differential conductance for $\alpha=2.05, \lambda=3.0$.
      Note that the first excitation now leads to PDC,
      cf.Fig.~\ref{fig:G_trap1}.
    }
  \end{center}
\end{figure}
\begin{figure}
  \begin{center}
    \includegraphics[scale=0.3,angle=-90]{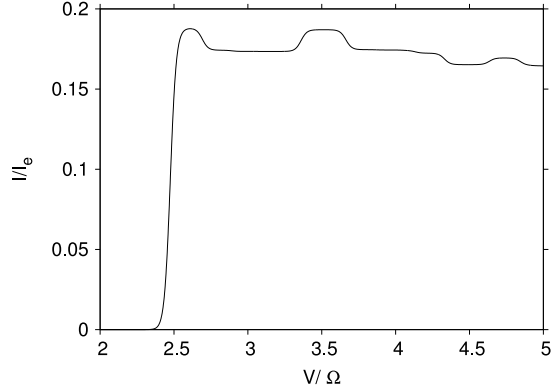}
    \caption{ \label{fig:I_trap2}
      $I(V)$ normalized to the electronic limit $I_e$
      for $\mu/\Omega=1.25$ in Fig.~\ref{fig:G_trap2}.
    }
  \end{center}
\end{figure}
\begin{figure}
  \begin{center}
    \includegraphics[scale=0.6,angle=-90]{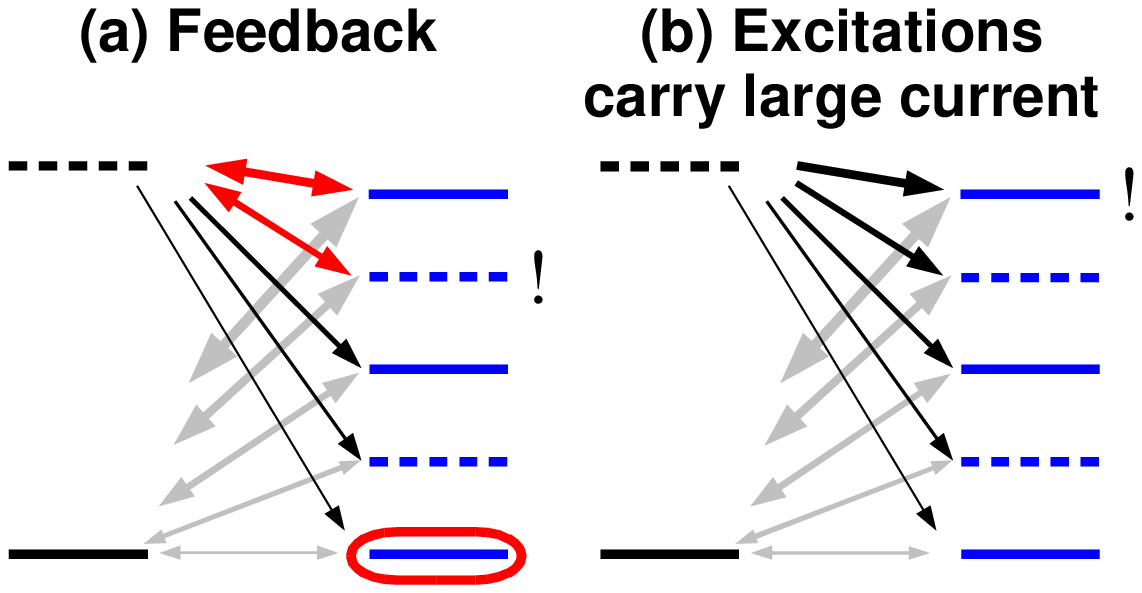}
    \caption{
      \label{fig:scheme-trap2}
      Transitions relevant to the NDC effects on the right in
      Fig.~\ref{fig:G_trap2}.
      (a) For most bias voltages transitions are switched on (red
      arrows) which eventually feed back  (black arrows) into the vibrational
      ground state $1_0$ .
      Escape from the highest states $[\alpha^2],[\alpha^2]-1,\ldots$
      with largest contributions to current becomes possible first,
      leading to the strongest NDC.
      (b) At the ground state transition line
      all the excitations $1_{m'},m'=1,\ldots,[\alpha^2]$ initially become
      occupied (gray arrows) and contribute to the a relatively large current
      without the feedback occuring.
    }
  \end{center}
\end{figure}
The strong asymmetry between the excitation spectra due to the
distortion results in the pronounced asymmetric conductance plot in
Fig.~\ref{fig:G_trap2}.
As one approaches the strong shift regime
the NDC at resonances $\mu_L=m'\omega_1,m'=1,\ldots, [\alpha^2]$
discussed below (Fig.~\ref{fig:G_par})
turns into PDC 
(cf. the first excitation in Fig.~\ref{fig:G_cohsup})
and becomes suppressed in intensity at low bias for $\lambda^2 > \alpha$
as seen in  Fig.~\ref{fig:G_trap2}.
This is also described by Eq.~(\ref{eq:Iplat}) (which holds for any
$\lambda$) since 
the FC-factors increase with $m',m$ for $\lambda^2 > \alpha$,
 cf. Fig.~\ref{fig:fc}(b).
For negative gate energy $\mu$ the current slowly increases, but for
positive $\mu$ the current surprisingly shows a sharp increase
followed by NDC, despite the strong shift.
Fig.~\ref{fig:scheme-trap2} explains how the postponement of the
classical trapping effect (similar to that discussed in
Sec.~\ref{sec:trap1}) allows for large current steps despite the
strongly shifted potentials.
This results in strong PDC excitations spaced by the
larger frequency $\omega_0$ and in between several pronounced NDC excitations
with smaller spacing $\omega_1$ (in contrast to Fig.~\ref{fig:G_trap1}
where the NDC spacing is $\omega_0-\omega_1$).
\section{Discussion
\label{sec:discuss}
}
We have found that non-equilibrium vibrational effects are enhanced in
molecular devices for which the effective potential for vibrations
is sensitive to the charge state of the device.
We modeled this by a change in the vibrational frequency in addition to a
shift of the potential minima.
In particular,
for weak distortion of the potential
the current was shown to map out sharp changes in the vibrational
distributions with bias voltage.
For sufficiently strong distortion of the potential interference
effects of the nuclear wave functions were show to strongly influence
the electron transport.
The coherent effects are also expected to occur for more detailed
models of the nuclear potentials since the requirements on the low-energy
vibrational excitations are rather basic.
The parity effect requires potentials which are
distorted and only slightly shifted.
The coherent suppression due to the vanishing overlap between an
vibrational excited- and ground-state corresponding to a different
charge on the molecule,
requires a moderate shift and distortion.
The precise conditions for the suppression will be different but
Fig.~\ref{fig:zeros} provides the qualitative picture.
These mechanism, together with possible weak dependence on current or
voltages offer interesting possibilities for controlling electron
transport in single molecule devices.
\appendix
\section{Franck-Condon factors for shifted and distorted potentials
 \label{app:fc}}
In this Appendix we give the expressions for the FC-factors for
potentials exhibiting both a relative shift and distortion.
The derivation can be done by straightforward algebra without recourse
to special functions (cf.~\cite{Hutchisson30,Manneback51,Siebrand66}).
We first note that the sign of the shift is irrelevant since the
each of the states in the overlap integral has a definite parity with
respect to spatial inversion relative to the minimum of its potential.
Using this we find that interchanging $\omega_{0}\leftrightarrow \omega_{1}$
(i.e. $\alpha=\sqrt{\omega_0/\omega_1} \rightarrow \alpha^{-1}$ is equivalent to charge
conjugation $N \rightarrow 1-N,N=0,1$ or $\mu \rightarrow -\mu$.
We can thus restrict ourselves to $\alpha>1$ as long as we discuss
both polarities of $\mu$.
For the calculation of $F_{m'm}$ it is convenient to normalize
the coordinate to ZPM of the potential in question,
$Q_N = x \sqrt{M \omega_N}$ and its conjugate $P_N$:
\begin{eqnarray}
  H_N &=& \frac{\omega_N}{2}
          \left[
            P_N^2 +
            \left( Q_N \pm \sqrt{\alpha^{\pm}/2} \lambda \right)^2
          \right]
   \label{eq:HN_norm}
 \end{eqnarray}
 where $N = 0, 1$ corresponds to~$\pm$ and the shift parameter is
 $\lambda = \sqrt{{M \Omega}/{2}} \delta x$
and the mean frequency $\Omega = \sqrt{\omega_0 \omega_1}$.
We obtain oscillator 1 from oscillator 0 by first applying a shift and then a
distortion:
 \begin{eqnarray*}
  Q_1 & = & \frac{Q_0}{\alpha} + \sqrt{2} \frac{\lambda}{\sqrt{\alpha}}
\end{eqnarray*}
We write the corresponding transformation of the
lowering operators, $b_N = \left( Q_N + i P_N \right) / \sqrt{2}$,
\begin{eqnarray*}
  b_1 & = & \frac{1}{2} \left( \frac{1}{\alpha} + \alpha \right) b_0 +
  \frac{1}{2} \left( \frac{1}{\alpha} - \alpha \right) b_0^{\dag} +
  \frac{\lambda}{\sqrt{\alpha}}
  ,
\end{eqnarray*}
as a unitary transformation $b_1 = e^{- S} b_0 e^S$, where
\begin{eqnarray*}
  S & = & \ln \alpha
  \left[
     \frac{1}{2} \left( b^2_0 - b_0^{\dag 2} \right)
     - \frac{\lambda}{\sqrt{\alpha} - \frac{1}{\sqrt{\alpha}}}
       \left( b_0 -  b_0^{\dag} \right)
   \right]
   .
\end{eqnarray*}
The exponential can be disentangled by the methods described in~\cite{Zhang90}:
$ e^{- S} = e^{L b^{\dag 2} + l b^{\dag}}
            e^{\left( \ln C \right)  \left( b^{\dag}b+ \frac{1}{2} \right)+ c}
            e^{R b^2 + r b}$
where the parameters in the exponential factors are
\begin{equation*}
  \left.
    \begin{array}{l}
      R \\
      L
    \end{array}
  \right\} =  \pm \frac{\alpha^2 - 1}{2 \left( \alpha^2 + 1 \right)},
  \left.
    \begin{array}{l}
      r \\
      l
    \end{array}
  \right\} =  \mp \frac{\lambda}{\alpha^{\pm 1/2}}  \frac{2 \alpha}{\alpha^2 +  1},
  C  =  \frac{2 \alpha}{\alpha^2 + 1},
  c  =  - \lambda^2 \frac{\alpha}{\alpha^2 + 1}
  .
\label{eq:disentpar}
\end{equation*}
The FC-factors
 $F_{m'm}=|{}_1\braket{m'}{m}_0|^2
         =|{}_0\bra{m} e^{-S} \ket{m'}_0|^2$
 are then directly found from the matrix elements
\begin{eqnarray*}
  {}_0\langle m| e^{S} |m' \rangle_{0} & = &
  \sum_{k = 0}^{\min \left\{ m, m' \right\}}
  \Theta_{m k} \left( L, l \right)
  \Theta_{m' k} \left( R, r \right) C^{k + \frac{1}{2}} e^c
  \\
  \Theta_{m k} \left( R, r \right) & = & \sqrt{\frac{m!}{k!}}
  \sum_{s = 0}^{\left[\frac{m - k}{2} \right]}
    \frac{R^s r^{m - k - 2 s}}{s! \left( m - k - 2 s \right) !}
    \\
    & = & 
    \sqrt{\frac{m!}{k!}}
    \left( - i r \right)^{\left( m - k \right)}
    H_{m - k}\left( i \frac{r}{2 R} \right)
    \left( \frac{R}{r^2} \right)^{\left[ \frac{m -k}{2} \right]}
    \label{eq:fcfull}
\end{eqnarray*}
For $\alpha = 1$ one obtains the well-known expression
 $
 F^{\alpha=1}_{m'm}
 =
 e^{-\lambda^2} \frac{m!}{{m'}!} \lambda^{2|m-m'|}
 \left[ L^{|m-m'|}_{m}(\lambda^2 )\right]^2
 $
where $L$ is the associated Laguerre-polynomial and $m<m'$
and $F_{m'm}=F_{mm'}$.
For $\lambda = 0$ the selection rule Eq.~(\ref{eq:parsel}) is easily
verified. The nonzero matrix elements, for instance, for the
special case $m=0,m'=2k$ are
$F_{2k0} = \sqrt{1-\xi^2} \frac{(2k-1)!!}{(2k)!!} \xi^{2k}$
where
 $(2k)!!=(2k)(2k-2)\cdots 2$
,
 $(2k-1)!!=(2k-1)(2k-3)\cdots 1$
and $\xi=\frac{\alpha^2 - 1}{\alpha^2 + 1}$.
The expression in terms of the Hermite polynomials
$H_n$ reduces to the known result for $m'=0,m \ne 0$~\cite{Manneback51,Siebrand66}.
\section{Classical features of the FC-factors
  \label{app:class}}
The large scale variations of the FC-factor $F_{m'm}$ in the $\left(
m, m' \right)$ plane and their  effects on the transport 
have a simple classical interpretation.
It is important to discuss these if one wants to identify quantum
effects of the nuclear motion.
The central point is that the FC-factor becomes exponentially suppressed unless
the nuclear motions in the effective potentials of the two charge states
are compatible i.e.
 the phase-space trajectories of the two motions intersect.
The boundary between classically forbidden and allowed regions in the
$(m,m')$-plane is found by requiring that the simultaneous equations
(cf. Eq.~(\ref{eq:HN}))
\begin{eqnarray}
  \frac{1}{\alpha} P^2 + \alpha \left( Q - \frac{\lambda}{\sqrt{2}}
  \right)^2  & = & 2 m
  \nonumber
  \\
  \alpha P^2 + \frac{1}{\alpha} \left( Q + \frac{\lambda}{\sqrt{2}}
  \right)^2  & = & 2 m'
  \label{eq:orbits}
\end{eqnarray}
have at least \textit{one} real valued solution for $(P,Q)$
(the vibrational energies are $E_0 = \omega_{0}m$ and $E_1 = \omega_{1} m'$).
Within the classically allowed region there may be regions
of different overall intensity related to the appearance of {\em additional}
solutions.
The intersections of the elliptic orbits determined by
Eq.~(\ref{eq:orbits}) are illustrated in Fig.~\ref{fig:fc}.
In the case of shifted potentials, $\lambda > 0, \alpha = 1$, $Q$ always has a
single real solution. Two real solutions for $P$ exist if
\begin{equation}
  m + m' \ge
  \frac{1}{2} \left[ \lambda^2 + \frac{1}{\lambda^2} \left( m - m'
    \right)^2 \right]
  \label{eq:parabola}
\end{equation}
This parabola in the $\left( m, m' \right)$ plane, tilted by an angle $\pi/4$
relative to $m'$ axis, is the so-called Condon-parabola~\cite{Herzberg} depicted in
Fig.~\ref{fig:fc}(a).
The condition~(\ref{eq:parabola}) is equivalent to demanding that the
classical turning points of the two motions are interspersed.
In the case of distorted potentials, $\lambda = 0, \alpha > 1,$ the
requirement is
\begin{equation}
  \frac{1}{\alpha^2} \le \frac{m'}{m} \le \alpha^2
  \label{eq:lines}
\end{equation}
and real solutions are always four in number.
The left inequality ensures that $Q$ is real. The corresponding lower
boundary line in Fig.~\ref{fig:fc}(c) is equivalent to requiring the
potential energies of the motions are equal at the maximal coordinate ($P=0$).
The right inequality ensures a real solution for $P$
and corresponds to equal kinetic energy at $Q=0$
(i.e. not at the classical turning point).
For the general case $\lambda > 0, \alpha > 1$ two real solution exist
for $P$ when $m, m'$ lie inside a Condon-parabola
\begin{equation}
 m \frac{1}{\alpha} + \alpha m' >
 \frac{1}{2} \left[
   \lambda^2 + \frac{1}{\lambda^2} \left( m \frac{1}{\alpha} - \alpha  m'
   \right)^2
 \right]
 \label{eq:parabola2}
\end{equation}
which is tilted by the angle $\varphi$ relative to $m$ axis, where
$\tan \varphi = {\alpha^{-2}} = {\omega_{1}}/{\omega_{0}}$
i.e. the parabola tilts towards the axis corresponding to
the highest frequency.
It touches the axes at $m=\lambda^2 \alpha$
 and $m'=\lambda^2 /\alpha$, respectively, corresponding to the
 elastic energies.
The solutions for $Q$ are always real in this region.
However, an additional two real solutions
for $Q$ occur between the parabolic boundary and below the line
\begin{equation}
 m' = \alpha^2 m + \frac{\alpha^3}{\alpha^4 - 1} \lambda^2,
 \label{eq:lines2}
\end{equation}
beyond its tangent point with the parabola, which is located at
$m = (\lambda^2   \alpha) /(   \alpha^4-1)^2,
 m' =(\lambda^2 / \alpha)/ (1-\alpha^{-4})^2$.
In Fig.~\ref{fig:fc}(b) one sees that in this region
the FC-factors are clearly enhanced.
The above classical expressions thus give an simple guide to the
large scale structure of the complicated exact expression~(\ref{eq:fcfull}).
For $\lambda \rightarrow 0$ the parabola~(\ref{eq:parabola2}) becomes
very narrow and reduces to a line through the origin line $m' / m = 1
/ \alpha^2$, i.e. we recover Eq.~(\ref{eq:lines}).
For $\alpha \rightarrow 1$ the region with 4 solutions
moves to infinity with the tangent point of the parabola
and we retain~(\ref{eq:parabola}).
\section{Vibrational distribution and numerical convergence
  \label{app:vibdist}
}
Much of the behavior of the occupations $P^N_m$ and the
current as function of the applied voltages may be
understood from a simple scheme which is readily extended to
situations with multiple competing orbitals~\cite{Nowack05}.
First, to determine which {\em direct} transitions between states
relevant one draws the region
 $\mu_{R} / \Omega <  m' / \alpha - m \alpha < \mu_{L}/\Omega$
(``bias window'') into the grayscale plot of the FC-factors.
For $m',m$ within this region transitions $1_{m'}\leftrightarrow 0_m$ are
{both} allowed
whereas above/below this region only
 $1_{m'} \rightarrow 0_m$ /
 $1_{m'} \leftarrow 0_m$ is
allowed by electrons leaving / entering  the molecule via {\em both} tunnel junctions.
Importantly, $m',m$ outside the classically allowed region can be
disregarded.
In a second step we have to determine which states will become { occupied}
significantly via {\em cascades} of tunnel processes 
(unless the relaxation is extremely fast, $\gamma \gg \Gamma$, in which
 case we can solve Eq.~(\ref{eq:dotP}) using the vibrational equilibrium
 ansatz~\cite{Braig03a,Mitra04b}).
These may in principle allow arbitrarily high states to be reached.
The FC-factors which satisfy sum rules (cf.~\ref{sec:model}) will
prevent the average vibrational numbers $\sum_m m P^N_m$ from 
increasing indefinitely with $V$.
The expressions for the boundary curves (\ref{eq:parabola2}) and
(\ref{eq:lines2}) now become helpful in estimating the number of
vibrational states required to solve the {\em transport} problem with
good accuracy.
The points of intersection $(m'_r,m_r)$ of the edges of the bias
window with the boundary curves can be explicitly found (e.g. for
$\alpha=1$ these points have a quadratic dependence on both $V$ and
$\mu$).
We may truncate the infinite set of master equations beyond the
cut-offs for on $m$ and $m'$ estimated as
\begin{equation}
  m_c    = {\rm max}_r \left\{ m_r(\mu,V)    \right\},
  {m'}_c = {\rm max}_r \left\{ {m'}_r(\mu,V) \right\}
  .
\end{equation}
Beyond these points the asymmetry between the FC-factors in the gain and
loss terms in the stationary master equation
increases with $m$ and $m'$. Therefore the {\em occupations} will
start to strongly decrease.
The convergence of the {\em current} requires more states to be taken into
account for strong shifts $\lambda^2 \alpha^{\pm} \gg 1$ 
even when the distribution is already converged.
In this case the exponential increase of the FC-factors and the strong
decrease of the occupations with $m$ tend to cancel out. 
For small shift the cut-offs $m_c,m'_c$ overestimate the number of
required states.
The distortion generally widens the classically allowed region
i.e. transitions with larger change $m'-m$ become more probable which
improves the convergence.
\bibliography{cite}
\bibliographystyle{unsrt}
\end{document}